\documentclass[10pt,journal,twocolumn,twoside]{IEEEtran} % formal style
\usepackage{graphicx}
\usepackage{epstopdf}
\usepackage{float}
\usepackage{algorithm,algorithmic}
\usepackage{array}
\usepackage{amsmath}
\usepackage{amssymb}
\usepackage{mdwmath}
\usepackage{mdwtab}
\usepackage{eqparbox}
\usepackage{flushend}
\usepackage{fixltx2e}
\usepackage{cases}
\usepackage{bm}
\usepackage{multirow}
\usepackage{psfrag}
\usepackage[usenames]{color}
\usepackage{mathtools}

\usepackage{fixmath}
\usepackage{mathdots}
\usepackage{latexsym}
\usepackage{amsmath,amssymb}

\usepackage{bm}
\usepackage{footmisc}

\usepackage[T1]{fontenc}% optional T1 font encoding
\usepackage{url}
\usepackage{cite}
\newcommand{\subparagraph}{}
\usepackage{titlesec}

\newcommand{\diag}{\mathop{\mathrm{diag}}}

\begin{document}

\title{Differential Reflecting Modulation for Reconfigurable Intelligent Surface Based Communications}
\author{
Shuaishuai~Guo,~\IEEEmembership{Member, IEEE,}
	Jia~Ye,~\IEEEmembership{Student Member, IEEE,}
Peng~Zhang,~\IEEEmembership{Member, IEEE,}  
Haixia~Zhang,~\IEEEmembership{Senior Member, IEEE,}  
        and   Mohamed-Slim Alouini,~\IEEEmembership{Fellow, IEEE}
\thanks{S. Guo and H. Zhang are with the Shandong Provincial Key Laboratory of Wireless Communication Technologies and School of Control Science and Engineering, Shandong University, Jinan 250061,
China (e-mail: shuaishuai\textunderscore{guo}@sdu.edu.cn;haixia.zhang@sdu.edu.cn).}
\thanks{J. Ye, M. -S. Alouini are with the Computer, Electrical and
Mathematical Science and Engineering Division,  King Abdullah University of Science and Technology, Thuwal, Saudi Arabia, 23955 (email: jia.ye@kaust.edu.sa; slim.alouini@kaust.edu.sa).}
\thanks{P. Zhang is with the School of Computer Engineering, Weifang University, Weifang 261061, China (e-mail: sduzhangp@163.com).}
}
%\markboth{Submitted to IEEE JOURNAL ON SELECTED AREAS IN COMMUNICATIONS}%
%{Guo \MakeLowercase{\textit{et al.}}:Generalized Beamspace Modulation Using Multiplexing: A Breakthrough in mmWave MIMO}
\maketitle
\begin{abstract}
Reconfigurable intelligent surface (RIS) based communications have emerged as a new paradigm. This letter proposes a differential reflecting modulation (DRM) scheme for RIS based communication systems. In DRM, information bits are jointly carried by the activation permutations of the reflecting patterns and the phases of the transmitted signals, leading to that DRM can work without any channel state information (CSI) at the transmitter, RIS or receiver. 
In other words, DRM can release the intricate and resource-consuming channel estimation in the transmission process. Simulation results show that the proposed DRM pays an acceptable SNR penalty compared to non-differential modulation with coherent detection.
\end{abstract}


\IEEEpeerreviewmaketitle
%\footnote{The work of S. Guo was partly supported by 
%China Postdoctoral Science Foundation under Grant 2017M622202 and by the
%National Natural Science Foundation of China under Grant 61471269.}

\section{Introduction}

\IEEEPARstart{R}{econfigurable} intelligent surface (RIS) is a key enabler for configuring favorable wireless communication environment \cite{di2019smart,Dang2020}. The research on RIS based communication systems has attracted a lot of interest. In most of RIS based communication systems, information is modulated onto the phases/magnitudes of the transmitted signals\cite{Ye2020,Qiao2020}, where RIS only plays the role of an assister helping improve the communication performance. Recently, there is a large body of literature designing RIS as an information modulator, e.g., \cite{tang2019wireless,tang2020wireless,tang2019programmable,basar2019transmission,guo2019reflecting}. To be more specific, Tang \emph{et al} designed  a RIS enabled phase modulation scheme in \cite{tang2019wireless} and radio frequency (RF) chain-free transmitters based on RIS in \cite{tang2020wireless} and \cite{tang2019programmable}. Basar \emph{et al} in \cite{basar2019transmission} proposed a RIS-enabled spatial modulation scheme, where information is not only modulated onto the phases/magnitudes of the transmitted signals but also onto the reflecting patterns.  Yan \emph{et al} proposed a passive beamforming and information transfer (PBIT), where RIS acts as a passive beamformer and also as an independent information modulator. \cite{guo2019reflecting} has made a comprehensive comparison among these RIS-enabled information transfers and proposed an optimized reflecting modulation (RM) scheme that outperforms all others.

So far, however, all existing schemes need to access the instantaneous/statistic channel state information (CSI) for detection. In RIS-based communication systems, the channel estimations involving transmitter-RIS,  RIS-receiver, and the direct transmitter-receiver  links are intricate tasks. This is because RIS units has no baseband signal precessing capabilities.  Although some works on solving the critical channel estimation problem have been done, e.g., \cite{Nadeem2020, Nadeem2020a}, the channel estimation itself still consumes a lot of time and resources. In this letter, we propose a differential reflecting modulation (DRM) scheme. In DRM, information bits are encoded onto the activation permutation order of the reflecting patterns and also the phases of the transmitted signals. Such designs enables differential detection bypassing the resource-consuming channel estimation. The detection complexity and transmission rate are analyzed. Simulations results show that the proposed DRM scheme pays acceptable  performance loss compared to  non-differential reflecting modulation (NDRM) with perfect CSI. Besides that, we show that DRM can achieve comparable or even better performance than NDRM with channel estimation errors. 
\section{System Model}
\begin{figure}[t]
  \centering
  \includegraphics[width=0.5\textwidth]{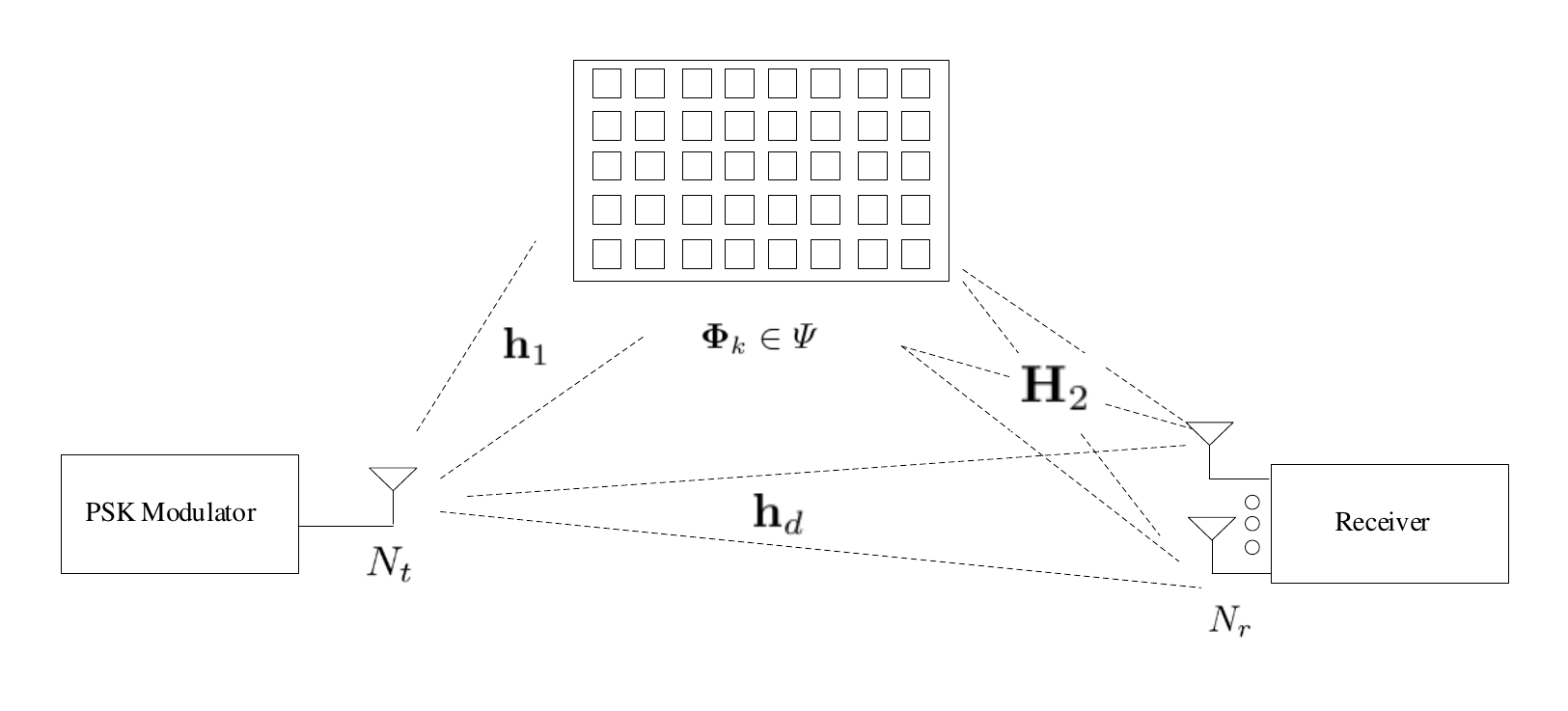}\\
 \caption{A RIS-assisted ($N_r$,$N$, $r$) SIMO communication system.}
  \label{System_Model}
\end{figure}
\begin{figure*}[t]
  \centering
  \includegraphics[width=0.75\textwidth]{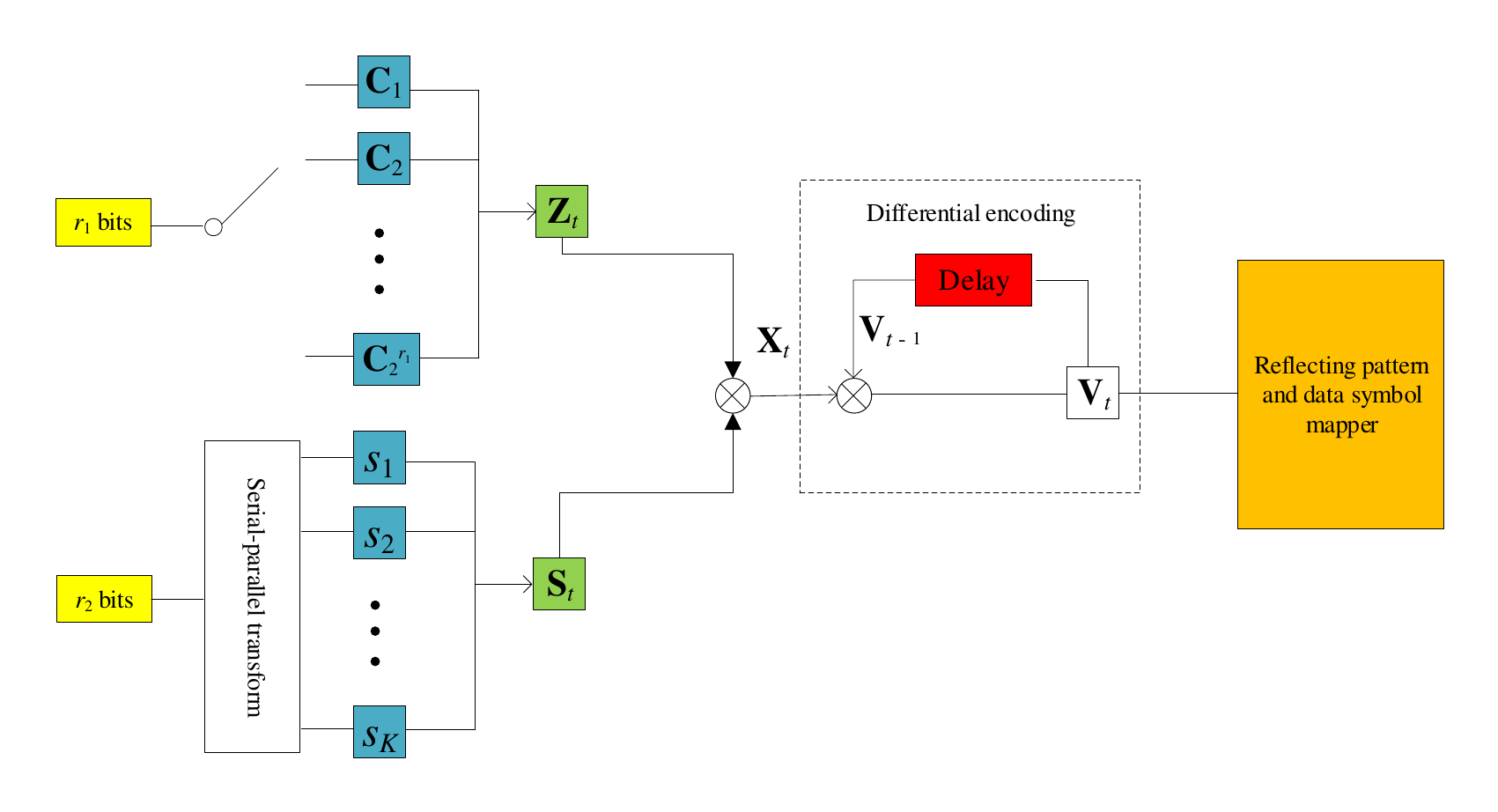}\\
 \caption{DRM encoding process.}
  \label{DRM_encoding}
\end{figure*}

In this letter, a RIS-assisted  $(N_r,N,r)$ single-input multiple-output (SIMO) communication system model as illustrated in Fig. \ref{System_Model}, where the transmitter is equipped with one antenna; $N_r$ denotes the numbers of receiver antennas; $N$ stands for the number of reflecting units on the surface; and $r$ stands for the total transmission rate. In the system, the transmitter sends $M$-ary phase shift-keying (PSK) symbols.  %Such system setups can model the communication between a power- and size- constrained transmitter  and a powerful receiver, such as the uplink communication in cellular systems.

Let $\mathbf{h}_1\in\mathbb{C}^{N\times 1}$, $\mathbf{H}_2\in\mathbb{C}^{N_r\times N}$, $\mathbf{h}_d\in\mathbb{C}^{N_r\times 1}$  denote the transmitter-RIS channel vector, the RIS-receiver channel matrix, and the direct link channel vector. We assume that the channels are quasi-static Rayleigh fading channels. It is worth noting that Rayleigh fading channels are chosen as an example to demonstrate the performance and the proposed DRM scheme are also applicable to other channel models. Moreover, we assume that a total of $K$ reflecting pattern candidates  are employed for transmission, which are included in a set $\Psi=\{\mathbf{\Phi}_1,\mathbf{\Phi}_2,\cdots \mathbf{\Phi}_K\}$. Each reflecting patten can be mathematically expressed as a diagonal matrix as $\mathbf{\Phi}_k\in\mathbb{C}^{N\times N}$. The diagonal elements of $\mathbf{\Phi}_k$ can be expressed as $(\mathbf{\Phi}_k)_{nn}=\beta_{n}^{(k)} \exp(j\theta_{n}^{(k)}), 1\leq n\leq N$, where $\beta_n^{(k)}=\{0,1\}$ indicates the ON/OFF states of the $n$th reflecting unit and $\theta_n^{(k)}$ represents the phase shift angle at the $n$-th reflecting unit when the $k$th reflecting pattern $\mathbf{\Phi}_k$ is activated.

\section{Differential Reflecting Modulation}
\subsection{Differential Encoding Scheme}
During the transmission, a frame is divided into blocks with each consisting of $K$ symbol time slots.  The rationale behind this dividing is that $K$ reflecting pattern candidates will be activated in a permutation order during a block, which will occupy $K$ symbol time slots.
At the $t$th block, a total of $r= \lfloor\log_2{ K!}\rfloor+K\log_2M$ bits are delivered. As shown in Fig. \ref{DRM_encoding},
the first $r_1=\lfloor\log_2{ K!}\rfloor$ bits are mapped to a $K\times K$ permutation matrix $\mathbf{Z}_t$. 
As is well known, there are $K!$ legitimate $K\times K$ permutation matrices. Here, we denote them as $\mathbf{C}_1,\mathbf{C}_2,\cdots, \mathbf{C}_{K!}$, receptively. In the mapping process, only $2^{r_1}$ matrices are chosen for the mapping process. To demonstrate the mapping process, an example with $r_1=2$ and $K=3$ is listed in Table I. In the demonstrated example, $4$ permutation matrices are chosen  from $6$ legitimate candidates for carrying $2$ bits.  It should be mentioned that different choices will result different performance. The rest $r_2=K\log_2M$ bits are mapped to $K$ $M$-PSK symbols $s_k~(1\leq k\leq K)$. By stacking all $K $ symbols in a vector $\mathbf{s}_t=[s_{1},s_{2},\cdots,s_{K}]$ and defining $\mathbf{S}_t=\diag{(\mathbf{s}_t)}\in\mathbb{C}^{K\times K}$, we introduce an information-carrying matrix $\mathbf{X}_t\in\mathbb{C}^{K\times K}$ given by
\begin{equation}
\mathbf{X}_t=\mathbf{Z}_t\mathbf{S}_t.
\end{equation}

Based on such an encoding mechanism, the transmitter and RIS can either jointly or separately delivery information. If the transmitter and RIS are connected by backhaul links and share the same information source, they can jointly transmit information of $r$ bits. Besides that, the transmitter and RIS can also separately delivery information. In this case, the transmitter and RIS does not need to know the information of each other. At each block, RIS sends $r_1$ bits of its own and the transmitter sends $r_2$ bits of its own. It is noteworthy that the RIS and the transmitter have to be synchronized in a wired or wireless way for both cases. The synchronization is beyond this scope of this letter and we assume that the transmitter and RIS are well synchronized.

Based on $\mathbf{X}_t$, a new matrix $\mathbf{V}_t\in \mathbb{C}^{K\times K}$ can be generated after the differential encoding as
\begin{equation}\label{eq11}
\mathbf{V}_t=\mathbf{V}_{t-1}\mathbf{X}_t,
\end{equation}
where $\mathbf{V}_{t-1}$ is the matrix generated in the former block and $\mathbf{V}_0\in\mathbb{C}^{K\times K}$ is  an identity matrix for initialization.
Based on the definition of $\mathbf{V}_t$, it is easily verified that  $\mathbf{V}_t$ is a multiplication of a permutation matrix and a diagonal matrix as
\begin{equation}\label{Vt}
\mathbf{V}_t=\tilde{\mathbf{Z}}_t\tilde{\mathbf{S}}_t,
\end{equation}
where $\tilde{\mathbf{Z}}_t\in\mathbb{C}^{K\times K}$ denotes a permutation matrix and $\tilde{\mathbf{S}}_t\in\mathbb{C}^{K\times K}$ represents a diagonal matrix whose symbols are  chosen from the $M$-PSK symbol set $\mathcal{S}_M$. This is because 
$\mathbf{V}_t=\mathbf{V}_0\mathbf{Z}_1\mathbf{S}_1\mathbf{Z}_2\mathbf{S}_2\cdots\mathbf{Z}_t\mathbf{S}_t$.
As shown in Fig. \ref{DRM_encoding}, next step is to activate the reflecting pattens and modulate the phases of transmit signals according to $\mathbf{V}_t$ during the $t$th block, which will be given in the following subsection.

 \begin{table}[t] %开始一个表格environment，表格的位置是h,here。
\centering
\caption{A Bit Encoding Example With $r_1=2$ and $K=3$} %显示表格的标题
\begin{tabular}{ | c | c |}%设置了每一列的宽度，强制转换。
\hline
\textbf{$r_1$ Bits} ($r_1=2$) & A $K\times K$ permutation matrix $\mathbf{Z}_t$ ($K=3$)\\ %用&来分隔单元格的内容 \\表示进入下一行
\hline
00 &$\mathbf{C}_1=\left[\begin{array}{ccc}
    1 & 0 & 0\\
    0 & 1 & 0\\
    0 & 0 & 1\\
\end{array}\right]$\\
\hline
01 & $\mathbf{C}_2=\left[\begin{array}{ccc}
    1 & 0 & 0\\
    0 & 0 & 1\\
    0 & 1 & 0\\
\end{array}\right]$\\
\hline
10& $\mathbf{C}_3=\left[\begin{array}{ccc}
    0 & 1 & 0\\
    1 & 0 & 0\\
    0 & 0 & 1\\
\end{array}\right]$\\
\hline
11& $\mathbf{C}_4=\left[\begin{array}{ccc}
    0 & 0 & 1\\
    1 & 0 & 0\\
    0 & 1 & 0\\
\end{array}\right]$\\
\hline
\end{tabular}
\end{table}

\subsection{Signal Transmission}
 Let $\mathbf{v}_k^{(t)}\in \mathbb{C}^{K\times 1}$ being $k$-th column vector of $\mathbf{V}_t$ and according to the form of $\mathbf{V}_t$ in (\ref{Vt}), $\mathbf{v}_k^{(t)}$ can be expressed as
\begin{equation}
 \mathbf{v}_k^{(t)}=\mathbf{e}_i^{(t)}v_k^{(t)},
\end{equation}
where  $\mathbf{e}_i^{(t)}\in \mathbb{C}^{K\times 1}$ represents the $i$th basis vector with $i$th elements being nonzero; and $v_k^{(t)}$ is the nonzero element of  $\mathbf{v}_k^{(t)}$ at the $i$th position, which is an $M$-PSK symbol. Then, according to $\mathbf{v}_k^{(t)}$, RIS activates the $i$th reflecting pattern $\mathbf{\Phi}_i$ and the transmitter sends the $M$-PSK symbol $v_k^{(t)}$ at the $k$-th slot of the $t$-th block.
Thus, the received signal $\mathbf{y}_k^{(t)}\in\mathbb{C}^{N_r\times 1}$ in the $k$-th slot of the $t$-th block can be written as\footnote{ In this letter, we consider the transmission is over quasi-static channels and assume that the time spread of multi-paths is not larger than a symbol time duration. }
\begin{equation}
\mathbf{y}_k^{(t)}=(\mathbf{h}_d+\mathbf{H}_{2}\mathbf{\Phi}_i\mathbf{h}_1)v_k^{(t)}+\mathbf{n}_k^{(t)},
\end{equation}
where $\mathbf{n}_k^{(t)}$ represents the complex Gaussian noise vector with zero mean and $\sigma^2\mathbf{I}_{N_r}$.

By introducing the following matrices
\begin{equation}
\tilde{\mathbf{H}}_d=[\overbrace{\mathbf{h}_d,\mathbf{h}_d,\cdots, \mathbf{h}_d}^{K}]\in\mathbb{C}^{N_r\times K},
\end{equation}
\begin{equation}
\tilde{\mathbf{H}}_2=[\overbrace{\mathbf{H}_2,\mathbf{H}_2,\cdots, \mathbf{H}_2}^{K}]\in\mathbb{C}^{N_r\times KN} ,
\end{equation}
\begin{equation}
\mathbf{Q}=
\begin{bmatrix}
    ~\mathbf{\Phi}_1 & \mathbf{0} &\cdots&\mathbf{0}\\
    ~\mathbf{0} & ~\mathbf{\Phi}_2& \cdots& \mathbf{0}~ \\
~\vdots& \vdots &  \ddots& \vdots~\\
    ~\mathbf{0} &\mathbf{0} &\cdots &\mathbf{\Phi}_{K}\\
\end{bmatrix}\in \mathbb{C}^{KN\times KN},
\end{equation} 
\begin{equation}
\tilde{\mathbf{H}}_1=\overbrace{
\begin{bmatrix}
    ~\mathbf{h}_1 & \mathbf{0} &\cdots&\mathbf{0}\\
    ~\mathbf{0} & ~\mathbf{h}_1& \cdots& \mathbf{0}~ \\
~\vdots& \vdots &  \ddots& \vdots~\\
    ~\mathbf{0} &\mathbf{0} &\cdots &\mathbf{h}_1\\
\end{bmatrix}}^{K}\in \mathbb{C}^{ KN\times K},
\end{equation} 
we rewrite the expression of $\mathbf{y}_k^{(t)}$ as
\begin{equation}
\mathbf{y}_k^{(t)}=(\tilde{\mathbf{H}}_d+\tilde{\mathbf{H}}_2\mathbf{Q}\tilde{\mathbf{H}}_1)\mathbf{v}_k^{(t)}+\mathbf{n}_k^{(t)}.
\end{equation}
By setting $\mathbf{H}=\tilde{\mathbf{H}}_d+\tilde{\mathbf{H}}_2\mathbf{Q}\tilde{\mathbf{H}}_1\in\mathbb{C}^{N_r\times K}$, the transmission can be expressed as
\begin{equation}
\mathbf{y}_k^{(t)}=\mathbf{H}\mathbf{v}_k^{(t)}+\mathbf{n}_k^{(t)},
\end{equation}
where $\mathbf{H}$ can be regarded as the equivalent channel matrix and $\mathbf{v}_k^{(t)}$ can be regarded as the equivalent transmit vector. Thus, during the $t$th block, the received signal matrix $\mathbf{Y}_t\in\mathbb{C}^{N_r\times K}$ can be expressed by 
\begin{equation}\label{eq10}
\mathbf{Y}_t=\mathbf{H}\mathbf{V}_{t}+\mathbf{N}_t,
\end{equation}
where $\mathbf{Y}_t=[\mathbf{y}_1^{(t)},\mathbf{y}_2^{(t)},\cdots, \mathbf{y}_K^{(t)}]\in\mathbb{C}^{N_r\times K}$ represents the received signal matrix, and $\mathbf{N}_t=[\mathbf{n}_1^{(t)},\mathbf{n}_2^{(t)},\cdots, \mathbf{n}_K^{(t)}]\in\mathbb{C}^{N_r\times K}$ is the complex Gaussian noise matrix.

\subsection{Detection Method}
%The advantage of the proposed DRM is that it can be detected without knowing any CSI at the transmitter, RIS or receiver. 
Substituting (\ref{eq11}) into (\ref{eq10}) yields
\begin{equation}
\begin{split}
\mathbf{Y}_t&=\mathbf{H}\mathbf{V}_{t-1}\mathbf{X}_t+\mathbf{N}_t\\
%&=(\mathbf{Y}_{t-1}-\mathbf{N}_{t-1})\mathbf{X}_t+\mathbf{N}_t\\
&=\mathbf{Y}_{t-1}\mathbf{X}_t-\mathbf{N}_{t-1}\mathbf{X}_t+\mathbf{N}_t.
\end{split}
\end{equation}
Thus, the optimal maximum-likelihood (ML) detector can be performed without knowing any CSI by 
\begin{equation}
\begin{split}
\hat{\mathbf{X}}_t&=\arg\min_{\mathbf{X}_t\in\mathcal{X}}||\mathbf{Y}_t-\mathbf{Y}_{t-1}\mathbf{X}_t||_{F}^2\\
%&=\arg\min_{\mathbf{X}_t\in\mathcal{X}} \textrm{tr}\left\{(\mathbf{Y}_t-\mathbf{Y}_{t-1}\mathbf{X}_t)^H(\mathbf{Y}_t-\mathbf{Y}_{t-1}\mathbf{X}_t)\right\}\\
&=\arg\max_{\mathbf{X}_t\in\mathcal{X}} \Re\left\{\rm{tr}(\mathbf{Y}_t^H\mathbf{Y}_{t-1}\mathbf{X}_t)\right\},
\end{split}
\end{equation}
where $\mathcal{X}$ is the set of all legitimate $\mathbf{X}_t$ and $|\mathcal{X}|=2^r$. Then, information bits can be decoded from $\hat{\mathbf{X}}_t$ by according to the mapping rule given in Subsection III-A.

\subsection{Transmission Rate and Complexity Analysis}
Considering the block for sending $\mathbf{V}_0$ does not contain information, the effective transmission rate of DRM can be written as
\begin{equation}
R=\frac{{(T-1)}r}{{T}K}=\frac{(T-1)\left(\lfloor\log_2 K!\rfloor+K\log_2M\right)}TK ,
\end{equation}
in bits per channel use (bpcu), where $T$ denotes the number of blocks in total. Based on the Stirling formula, i.e., $K!\approx \sqrt{2\pi K}(K/e)^K$ \cite{Guo2016},
the transmission rate is rewritten as
\begin{equation}
R\approx \log_2 M+\lfloor\log_2 \sqrt{2\pi K}+K\log_2(K/e)\rfloor/K,
\end{equation}
when $T$ is sufficiently large.
It can be checked that the transmission rate increases as $K$ increases, but the increase rate is not fast. The detection computational complexity can be analyzed to be 
$\mathcal{C}_1=2^{r}(K^2N_r+K^3) ~\textrm{multiplications}$, 
 since it needs to compute $\mathbf{Y}_t^H\mathbf{Y}_{t-1}\mathbf{X}_t$ by $2^r$ times and each computation requires $K^2N_r+K^3$ multiplications. Recall that $r=\lfloor\log_2 K!\rfloor+K\log_2M$, the computational complexity can be expressed as $\mathcal{C}_1=(2^{\lfloor\log_2 K!\rfloor}+M^K)(K^2N_r+K^3)~\textrm{multiplications}$.  Based on the expression of $\mathcal{C}_1$, it is found that the detection complexity increases much greatly as $K$ increases. 

\subsection{Reflecting Pattern Selection}
Above analysis indicates a good choice is to choose a small number of reflecting patterns for DRM, which can enjoy low-complexity detection. Then, how to perform reflecting pattern selection in a finite set for transmission arouses our interest. Since CSI is neither known by the transceivers nor by the RIS, we propose an optimization criterion to maximize the minimum mutual Euclidean distances, which is defined as $
d_{\min}=\min_{\mathbf{\Phi}_i,\mathbf{\Phi}_{i'}\in\Psi,s_k,s_{k'}\in\mathcal{S}_M\atop{\mathbf{\Phi}_is_k\neq\mathbf{\Phi}_{i'} s_{k'}}}||\mathbf{\Phi}_is_k-\mathbf{\Phi}_{i'}s_{k'}||_2 $.
 In this letter, we adopt the stepwise depletion algorithm proposed in \cite{guo2019reflecting}.  The detailed procedure of the stepwise depletion algorithm  can be found in \cite{guo2019reflecting} and we omit it for brevity.

\subsection{Impact of Channel Variation on DRM}
When the channel varies, the proposed DRM will inevitably lead to detection errors like other differential modulation schemes. After the variation, i.e., when the channel becomes static again, the proposed DRM will work well again.
To alleviate the impact, the transceiver can adopt bit interlevers/deinterleavers and error correcting codecs to correct the detection errors. Specifically, the bit interlevers/deinterleavers can spread the bit errors during the channel variation to multiple blocks resulting into a few of error bits per block, which can be corrected by error correction codecs.

\section{Simulations}

\begin{figure}[t]
  \centering
  \includegraphics[width=0.45\textwidth]{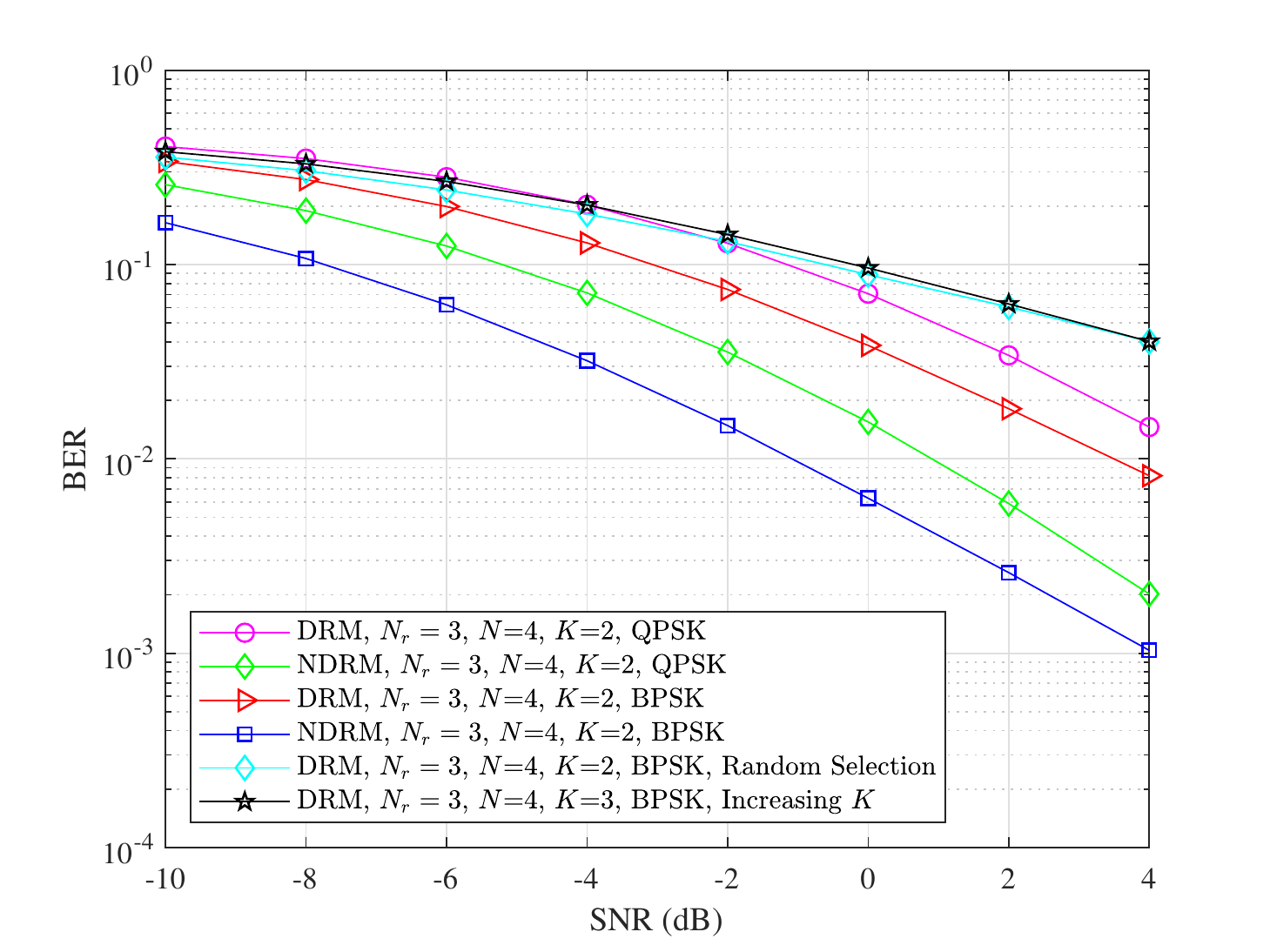}\\
 \caption{Performance comparison in RIS-assisted communication system with perfect CSI.}
  \label{result2}
\end{figure}

To show the performance of the proposed DRM scheme in $(N_r,N,r)$ RIS-based SIMO communication systems, we  compare DRM with NDRM.  In NDRM, $\mathbf{X}_t$ is an transmission matrix and the received signal matrix $\hat{\mathbf{Y}}_t=\mathbf{H}\mathbf{X}_{t}+\mathbf{N}_t$. Perfect CSI $\mathbf{H}=\tilde{\mathbf{H}}_d+\tilde{\mathbf{H}}_2\mathbf{Q}\tilde{\mathbf{H}}_1$ (i.e., perfect $\mathbf{H}_2$, $\mathbf{h}_d$ and $\mathbf{h}_1$) is adopted for coherent detection. The rationale behind choosing the comparison is that the transmission rate and detection complexity of both schemes are the same. In the comparison, we assume RIS is with $N=4$ units with each unit being $1$-bit encoded. That is, $(\mathbf{\Phi}_k)_{nn}\in\{-1,1\}$. Based on the assumption, we have $2^N=16$ legitimate reflecting patterns for DRM and NDRM transmission. We chose two  (i.e., $K=2$)  of them using  the stepwise depletion algorithm \cite{guo2019reflecting}. The simulation results are depicted in Fig. \ref{result2}. As the figure shows, DRM is less comparable to NDRM by $3-5$ dB under the given simulation setups. And in the simulations, the SNR is $\rho=\mathbb{E}\{|v_k|^2\}/\sigma^2=1/\sigma_n^2$. For more comparison, we include the BER of DRM with an increasing $K$ (from $K=2$ to $K=3$).  Simulation results show that the error performance of DRM decreases as $K$ increases.  In addition, we also demonstrate the superiority of the proposed reflecting pattern selection over a random selection in Fig. 3.

\begin{figure}[t]
  \centering
  \includegraphics[width=0.45\textwidth]{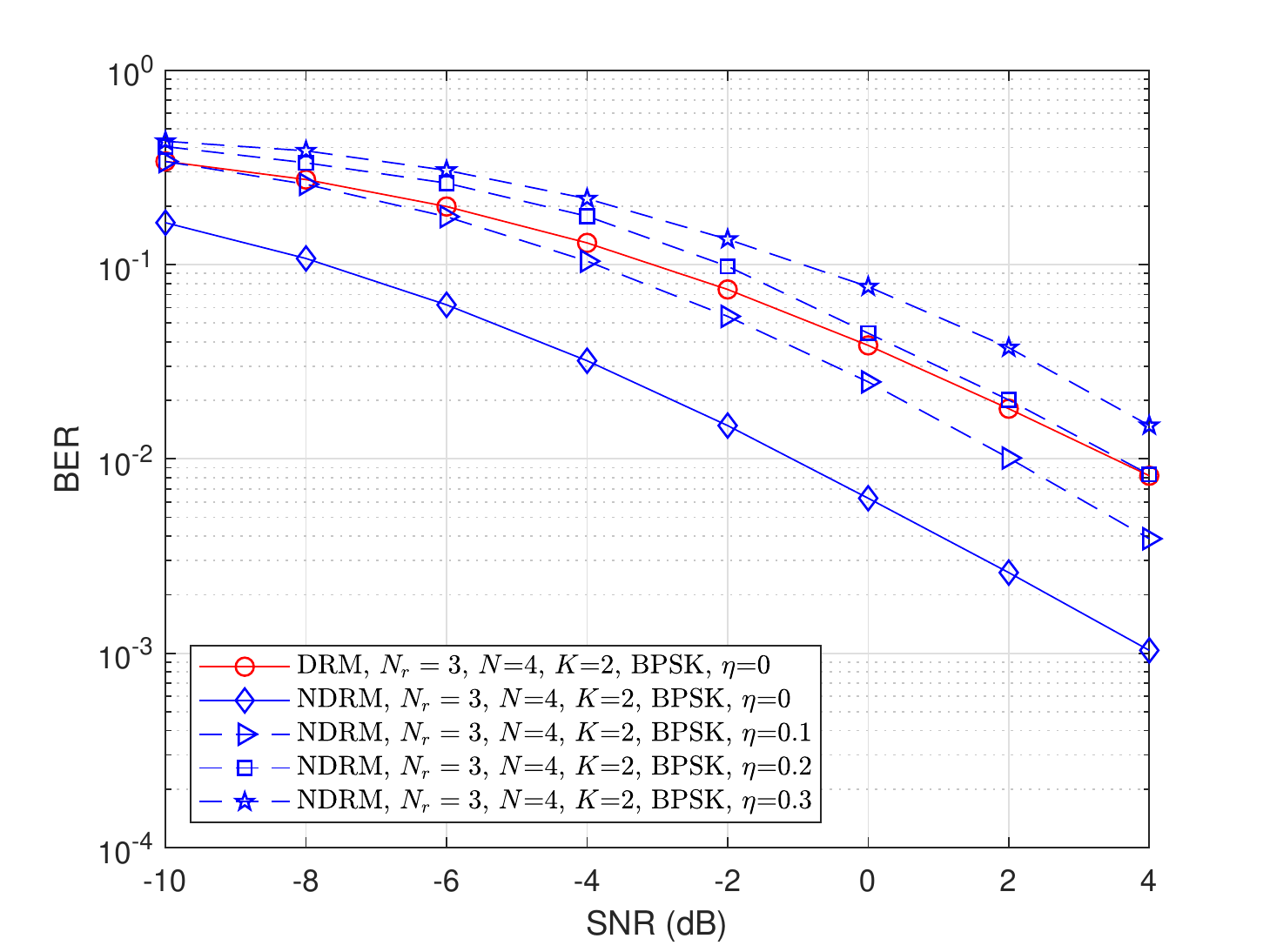}\\
 \caption{Performance comparison in RIS-assisted communication systems with imperfect CSI.}
  \label{result3}
\end{figure}
It is worth noting that DRM can work without CSI while NDRM have to spend much resource on channel estimation. Due to the fact that perfect CSI is typically not available, we further compare DRM and NDRM with imperfect CSI, where the imperfect CSI can be expressed by $\mathbf{H}^{\textrm{im}}_2=\mathbf{H}_2+\mathbf{H}_2^{\textrm{e}}$, $\mathbf{h}^{\textrm{im}}_d=\mathbf{h}_d+\mathbf{h}_d^{\textrm{e}}$ and $\mathbf{h}^{\textrm{im}}_1=\mathbf{h}_1+\mathbf{h}_1^{\textrm{e}}$. In the error model,  $\mathbf{H}_2^{\textrm{e}}$, $\mathbf{h}_d^{\textrm{e}}$ and $\mathbf{h}_1^{\textrm{e}}$ represent the error terms, with each elements in the matrix or vectors following a complex Gaussian distribution with zero mean and covariance $\sigma_e^2=\eta\sigma^2$ \cite{guo2019reflecting}, where $\eta$ is the positive proportional coefficient that is related to the number of pilots, the power and the employed algorithms for channel estimation. Simulation results are illustrated in Fig. \ref{result3}, which demonstrate that the performance of NDRM reduces much as CSI errors increase. The difference between the performance of DRM and that of NDRM becomes small when $\eta=0.1$. When $\eta=0.2$ and $\eta=0.3$, DRM can outperform NDRM with coherent detection in the depicted SNR regime.

\section{Conclusion}
In this letter, a differential modulation scheme named as DRM was proposed  for RIS-based communication systems. In DRM, the information bits are jointly encoded onto the permutation order of the reflecting patterns and the phases of the transmit signals, which can be detected without knowing the channel state information. Simulation results showed that the difference between the proposed DRM and NDRM with coherent detection is acceptable, especially when there are inevitable channel estimation errors.


\begin{thebibliography}{10}
\providecommand{\url}[1]{#1}
\csname url@samestyle\endcsname
\providecommand{\newblock}{\relax}
\providecommand{\bibinfo}[2]{#2}
\providecommand{\BIBentrySTDinterwordspacing}{\spaceskip=0pt\relax}
\providecommand{\BIBentryALTinterwordstretchfactor}{4}
\providecommand{\BIBentryALTinterwordspacing}{\spaceskip=\fontdimen2\font plus
\BIBentryALTinterwordstretchfactor\fontdimen3\font minus
  \fontdimen4\font\relax}
\providecommand{\BIBforeignlanguage}[2]{{%
\expandafter\ifx\csname l@#1\endcsname\relax
\typeout{** WARNING: IEEEtran.bst: No hyphenation pattern has been}%
\typeout{** loaded for the language `#1'. Using the pattern for}%
\typeout{** the default language instead.}%
\else
\language=\csname l@#1\endcsname
\fi
#2}}
\providecommand{\BIBdecl}{\relax}
\BIBdecl

\bibitem{di2019smart}
M.~Di~Renzo, M.~Debbah, D.-T. Phan-Huy, A.~Zappone, M.-S. Alouini, C.~Yuen,
  V.~Sciancalepore, G.~C. Alexandropoulos, J.~Hoydis, H.~Gacanin \emph{et~al.},
  ``Smart radio environments empowered by reconfigurable ai meta-surfaces: an
  idea whose time has come,'' \emph{EURASIP Journal on Wireless Communications
  and Networking}, vol. 2019, no.~1, pp. 1--20, May 2019.

\bibitem{Dang2020}
S.~Dang, O.~Amin, B.~Shihada, and M.-S. Alouini, ``What should {6G} be?''
  \emph{Nature Electronics}, vol.~3, no.~1, pp. 20--29, Jan. 2020.

\bibitem{Ye2020}
J.~{Ye}, S.~{Guo}, and M.~{Alouini}, ``Joint reflecting and precoding designs
  for {SER} minimization in reconfigurable intelligent surfaces assisted {MIMO}
  systems,'' \emph{IEEE Trans. Wireless Commun.}, early access, 2020.

\bibitem{Qiao2020}
J.~{Qiao} and M.~{Alouini}, ``Secure transmission for intelligent reflecting
  surface-assisted mmwave and terahertz systems,'' \emph{IEEE Wireless
  Communications Letters}, early access, 2020.

\bibitem{tang2019wireless}
W.~Tang, X.~Li, J.~Y. Dai, S.~Jin, Y.~Zeng, Q.~Cheng, and T.~J. Cui, ``Wireless
  communications with programmable metasurface: Transceiver design and
  experimental results,'' \emph{China Communications}, vol.~16, no.~5, pp.
  46--61, May 2019.

\bibitem{tang2020wireless}
W.~Tang, M.~Z. Chen, J.~Y. Dai, Y.~Zeng, X.~Zhao, S.~Jin, Q.~Cheng, and T.~J.
  Cui, ``Wireless communications with programmable metasurface: New paradigms,
  opportunities, and challenges on transceiver design,'' \emph{{IEEE} Wireless
  Commun. Mag.}, vol.~27, no.~2, pp. 180--187, Apr. 2020.

\bibitem{tang2019programmable}
W.~Tang, J.~Y. Dai, M.~Chen, X.~Li, Q.~Cheng, S.~Jin, K.-K. Wong, and T.~J.
  Cui, ``Programmable metasurface-based {RF} chain-free {8PSK} wireless
  transmitter,'' \emph{Electronics Letters}, vol.~55, no.~7, pp. 417--420, Apr.
  2019.

\bibitem{basar2019transmission}
E.~Basar, ``Transmission through large intelligent surfaces: A new frontier in
  wireless communications,'' in \emph{2019 European Conference on Networks and
  Communications (EuCNC)}.\hskip 1em plus 0.5em minus 0.4em\relax IEEE, 2019,
  pp. 112--117.

\bibitem{guo2019reflecting}
S.~Guo, S.~Lv, H.~Zhang, J.~Ye, and P.~Zhang, ``Reflecting modulation,''
  \emph{{IEEE} J. Sel. Areas Commun.}, early access, 2020.

\bibitem{Nadeem2020}
Q.~{Nadeem}, A.~{Kammoun}, A.~{Chaaban}, M.~{Debbah}, and M.~{Alouini},
  ``Asymptotic max-min {SINR} analysis of reconfigurable intelligent surface
  assisted {MISO} systems,'' \emph{{IEEE} Trans. Wireless Commun.}, early
  access, 2020.

\bibitem{Nadeem2020a}
Q.~{Nadeem}, H.~{Alwazani}, A.~{Kammoun}, A.~{Chaaban}, M.~{Debbah}, and
  M.~{Alouini}, ``Intelligent reflecting surface-assisted multi-user {MISO}
  communication: Channel estimation and beamforming design,'' \emph{IEEE Open
  Journal of the Communications Society}, vol.~1, pp. 661--680, May 2020.

\bibitem{Guo2016}
S.~{Guo}, H.~{Zhang}, P.~{Zhang}, and D.~{Yuan}, ``Link-adaptive mapper designs
  for space-shift-keying-modulated {MIMO} systems,'' \emph{IEEE Trans. Veh.
  Technol.}, vol.~65, no.~10, pp. 8087--8100, Oct. 2016.

\end{thebibliography}
\end{document}